\PassOptionsToPackage{table,dvipsnames}{xcolor}
\documentclass[conference]{IEEEtran}
\IEEEoverridecommandlockouts
\usepackage{xcolor}
\usepackage[table]{xcolor}
\usepackage{cite}
\usepackage{amsmath,amssymb,amsfonts}
\usepackage{graphicx}
\usepackage{textcomp}
\usepackage{algorithm}
\usepackage{algpseudocode}
\usepackage{booktabs} 
\usepackage{tikz}
\usetikzlibrary{shapes.geometric, arrows.meta, positioning}
\def\BibTeX{{\rm B\kern-.05em{\sc i\kern-.025em b}\kern-.08em
    T\kern-.1667em\lower.7ex\hbox{E}\kern-.125emX}}
\begin{document}

\title{CHILDES-Aligned: A Curated Children's Speech Dataset via Multi-Model Timestamp Ensembling \\
}

\author{
\IEEEauthorblockN{
Haolong Zheng\textsuperscript{1,\dag},
Yuanzhuo Hu\textsuperscript{2,\dag},
Xinyu Liang\textsuperscript{2},
Vishal Sunder\textsuperscript{3},
Dancheng Liu\textsuperscript{4},\\
Jinjun Xiong\textsuperscript{4},
Samuel Thomas\textsuperscript{3},
Brian Kingsbury\textsuperscript{3},
Zhizheng Wu\textsuperscript{2},
Mark A. Hasegawa-Johnson\textsuperscript{1,*}
\thanks{\textsuperscript{\dag}These authors contributed equally.}
\thanks{\textsuperscript{*}Corresponding author. E-mail: jhasegaw@illinois.edu}
}
\IEEEauthorblockA{
\textsuperscript{1}University of Illinois Urbana-Champaign \quad
\textsuperscript{2}The Chinese University of Hong Kong, Shenzhen \\
\textsuperscript{3}IBM Research \quad
\textsuperscript{4}University at Buffalo
}
}

\maketitle

\begin{figure*}[htbp]
    \centering
    \includegraphics[width=1.0\textwidth]{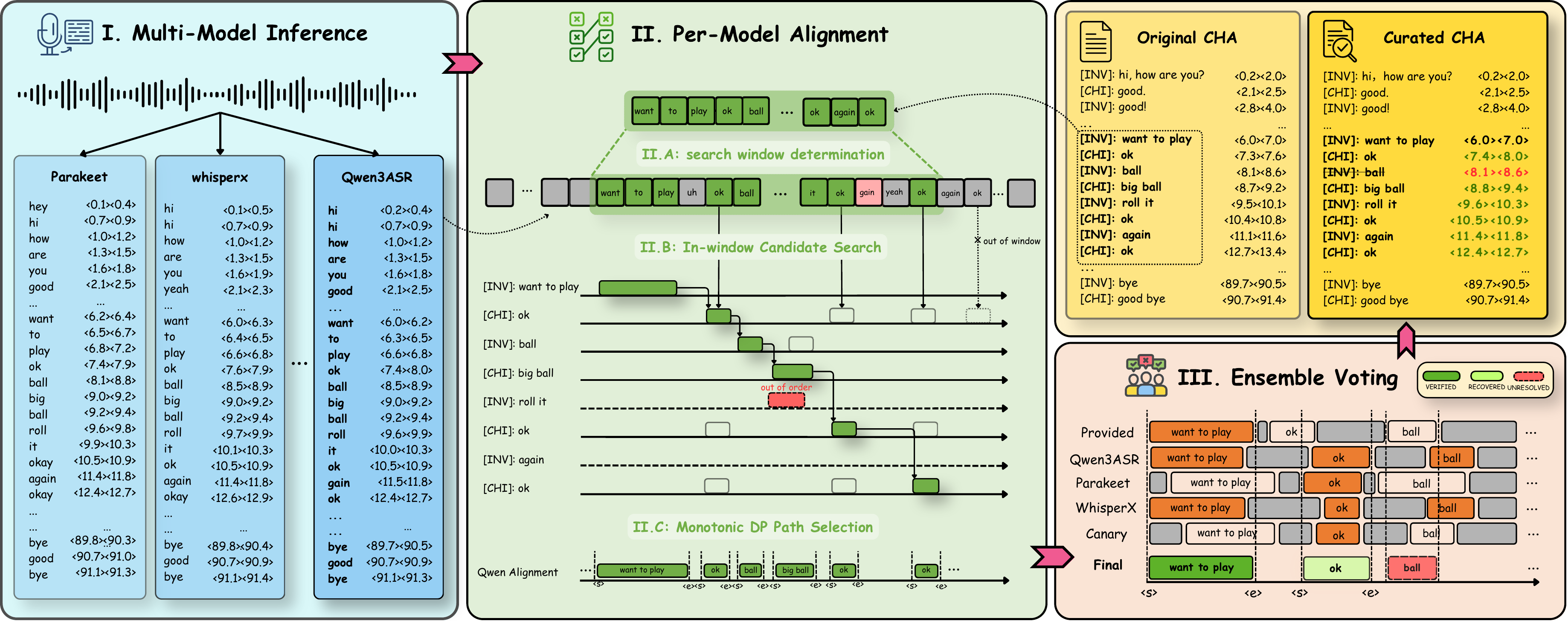}
    \caption{\textbf{Overview of the \textsc{Beacon} pipeline}}
    \label{fig:pipeline}
\end{figure*}


\begin{abstract}
CHILDES is a large-scale child speech corpus containing long-form recordings of naturalistic child-adult interactions, making it a valuable resource for studying child speech and language development. However, utterance-level timestamps provided in this corpus are often noisy, incomplete, or misaligned with the audio. As a result, utterances cannot always be reliably localized within long recordings, which limits the direct use of these data for training and evaluating speech models. In this work, we propose \textsc{Beacon} (\textbf{B}oundary \textbf{E}stimation via \textbf{A}lignment \textbf{CON}sensus)\footnote{Code and curated data: \texttt{https://github.com/MagicLuke/BEACON}}, an ensemble timestamp-curation framework that refines utterance-level timestamps by aggregating knowledge from multiple off-the-shelf ASR models. Specifically, each model's word-level timestamp predictions are first aligned to provided human transcripts, and the final utterance time boundaries are determined by a consensus voting strategy. The framework is corpus-agnostic and applies to any long-form recording paired with a trusted transcript whose timestamps are unreliable or missing, offering a general recipe for timestamp curation. Leveraging this pipeline, we curate and release a 413-hour general-purpose child-speech dataset with corrected utterance-level timestamps, together with a 283-hour quality-controlled subset for ASR training. Fine-tuning on this subset yields up to an average 19.5\% relative WER reduction on four out-of-domain child-speech benchmarks.

\end{abstract}

\begin{IEEEkeywords}
Child speech, Automatic Speech Recognition
\end{IEEEkeywords}

\section{Introduction}

Despite rapid progress in automatic speech recognition (ASR) and speech foundation models, child speech remains poorly supported\cite{fan2024benchmarking,bhardwaj2022automatic}. These systems are typically trained on adult speech\cite{shi2026qwen3asr,parakeet-canary, radford2023robust} and their performance often degrades when applied to children due to differences in acoustic and linguistic features which vary enormously with age and speaker\cite{ghai2010exploring,gerosa2009review,shivakumar2020transfer,koenig2008speech,koenig2008stop,lee1997analysis,lee1999acoustics,vorperian2007vowel}. Adaptation techniques for children’s ASR are still limited by the scarcity of high-quality labeled child speech data\cite{11461434,zheng2025ticl+,zheng2026sicl,zheng2026fsa,ying2025benchmarking,li2026age,jain2023adaptation,attia2024kid,liu2024sparsely,wang2026mind,fan2022draft, fan2022towards, jain2023wav2vec2}. As a result, current speech models often struggle to capture the full target distribution of child speech, reducing their robustness in real-world child-centered settings.

CHILDES\cite{macwhinney2000childes} provides a large collection of naturalistic child-adult interaction recordings, together with rich linguistic annotations and human transcripts. However, many recordings are long-form conversational sessions, and their utterance-level timestamps are often noisy, incomplete, or misaligned with the audio. This hinders the direct use of CHILDES for training and evaluating speech models.

Forced alignment is the natural tool here, since the transcript is known, but standard aligners~\cite{MFA,pratap2024mms} and long-form ASR~\cite{radford2023robust} degrade on long, noisy conversational audio, drifting and hallucinating boundaries across overlaps, disfluencies, and non-speech regions. The standard remedy is to first segment the recording and align each piece, using methods like recursive or CTC anchoring~\cite{moreno1998recursive,kurzinger2020ctcsegmentation,stan2016alisa,doras2023linearmemory} or by voice-activity detection~\cite{bain2023whisperx,FASA}. However, such re-segmentation assumes clean, single-stream transcripts. It also discards the speaker-attributed utterance structure and its human annotation, precisely the features that make CHILDES valuable.

Our contributions are as follows:

\begin{itemize}
  \item \textbf{A general long-form alignment pipeline.} We develop \textsc{Beacon}, a pipeline that recovers utterance-level timestamps from long-form audio paired with a trusted transcript, with or without preexisting coarse timestamps: it aligns the transcript against each ASR system's predictions to obtain a per-model timestamp, then aggregates these into a single consensus estimate (Section~\ref{sec:method}).
  \item \textbf{Curated CHILDES that preserves the original labels.} Applying the pipeline to English CHILDES, we correct its utterance timestamps without altering the original human transcripts and annotations, and release a lossless general-purpose version (over $400$ hours, raw CHAT text) with per-clip quality metadata (Section~\ref{sec:general}).
  \item \textbf{An ASR-ready subset, shown useful.} We derive a stricter subset ($283$ hours) with added quality control: transcripts are verbatim-normalized and every clip must pass an independent ASR agreement check. We validate it by fine-tuning ASR models on it and evaluating on out-of-domain child-speech benchmarks (Section~\ref{sec:asr}).
\end{itemize}

\section{Related Work}

High-quality labelled English child speech corpora remain scarce. CMU Kids\cite{CMU-Kids} provides 9 hours of read-aloud sentences. The OGI Kids' Speech corpus\cite{OGI} contains around 75 hours of recordings, mostly of read sentences, isolated words, or digit strings\cite{shahinmulti}. TIDIGITS\cite{leonard1993tidigits} includes digit strings spoken by children, while the Redmond Sentence Recall corpus\cite{RSR} provides recordings of 16 fixed sentences. Conversational and narrative collections are also limited in scale. MyST\cite{MyST} is relatively larger, comprising 393 hours of data with roughly 197 hours transcribed\cite{jain2023wav2vec2}. By contrast, CHILDES\cite{macwhinney2000childes} aggregates far more naturalistic and age‑diverse data from numerous independent corpora; however, its long‑form recordings often come with noisy timestamps, making them difficult to use for ASR training and evaluation. Also, to the best of our knowledge, many corpora are private and not publicly accessible\cite{hagen2003children, demuth2006word,russell2006pf, batliner2005pf_star, kazemzadeh2005tball}. All of these make our large-scale, high-quality labeled and publicly released dataset more valuable.

Curating CHILDES into usable clips requires aligning each CHAT transcript to its
long-form recording. Traditional forced aligners~\cite{Kaldi,MFA} are hard to
apply here: they assume the transcript closely matches clean, segmented audio
and degrade sharply on the long, noisy conversational recordings. Recent tools instead
build on ASR systems that emit word-level timestamps. BatchAlign2~\cite{BatchAlign2},
the official TalkBank toolkit, force-aligns the known CHAT transcript to the audio
with either an MMS/wav2vec2 CTC backend or a Whisper forced-alignment backend.
FASA~\cite{FASA} independently aligns each
WhisperX-predicted sentence to its best edit-distance span in the reference, keeping only the high-confidence matches and discarding the rest. In both tools, when the underlying model hallucinates or mistranscribes on the recordings, the error propagates uncorrected.
Aligning imperfect transcripts to long recordings has a long lineage, from iterative text-to-speech alignment robust to transcription errors~\cite{katsamanis2011sailalign} to transcript-biased recognition with confidence scoring~\cite{elmahdy2014alignment}. Our key idea is to pair such large-corpus text-to-text alignment with multi-recognizer consensus voting~\cite{fiscus1997rover,dietterich2000ensemble}, a long-standing ASR technique not previously combined with it, averaging out the model-specific errors a single aligner inherits.

\section{Methodology}
\label{sec:method}

\label{sec:overview}

We consider the general problem of recovering utterance-level timestamps for a long-form audio recording that is paired with a trusted reference transcript. We propose \textsc{Beacon}, a fully automatic algorithm that recovers a reliable onset and offset for each utterance, optionally correcting any preexisting coarse timestamps. Our method, has three steps (Figure~\ref{fig:pipeline}). \textbf{(1)~Multi-model inference}
(Section~\ref{sec:multiasr}): several architecturally diverse ASR systems transcribe the recording and produce word-level timestamp streams. \textbf{(2)~Per-model alignment}
(Section~\ref{sec:alignment}): each word stream is aligned independently to the reference utterance sequence, assigning each utterance a candidate time span or marking it as missing; this step consists of search-window determination, in-window candidate search, and monotonic dynamic-programming (DP) path selection. \textbf{(3)~Ensemble voting}
(Section~\ref{sec:ensemble}): the candidate spans from different models are combined by a consensus rule to produce a final utterance-level timestamp.

\subsection{Step 1: Multi-model inference}
\label{sec:multiasr}

We use four ASR systems with diverse architectures and timestamping mechanisms, so that consensus can reduce model-specific bias in utterance localization: Parakeet\footnote{\texttt{parakeet-tdt-0.6b-v3}}~\cite{xu2023efficient}, a token-and-duration transducer; Canary\footnote{\texttt{canary-1b-v2}}~\cite{parakeet-canary}, an attention encoder--decoder with CTC-aligned word timestamps; WhisperX~\cite{bain2023whisperx}, which applies VAD, Whisper transcription, then wav2vec2 forced alignment; and Qwen3-ASR\footnote{\texttt{Qwen3-ASR-1.7B}, \texttt{Qwen3-ForcedAligner-0.6B}}~\cite{shi2026qwen3asr}, whose transcript is timestamped by its LLM-based forced aligner. Each system $m$ is run with its standard recipe and emits a word stream $W^{(m)} = \bigl(w^{(m)}_1, \dots, w^{(m)}_{M_m}\bigr)$, where each word carries its start and end times in seconds.

\subsection{Step 2: Per-model alignment}

\label{sec:alignment}

This step prepares the inputs for consensus voting:
for a fixed model $m$, let its timestamped word stream be $W^{(m)}$ and let the
reference be a sequence of utterances $U = (u_1, \dots, u_N)$, where each
$u_i$ is a normalized token sequence with $|u_i|$ tokens. The alignment assigns
each utterance $u_i$ either a contiguous span $[a_i, b_i] \subseteq
\{1,\dots,M_m\}$ of model words from $W^{(m)}$, whose boundary timestamps define the
recovered utterance onset and offset (the time interval voted on in Step~3), or the label \textsc{Missing} when model
$m$ does not reliably cover the utterance.

The alignment has three stages. First, \textbf{search-window determination} localizes each utterance to a short region of the model word stream. Second, \textbf{in-window candidate search} proposes high-quality span candidates for each utterance. Finally, \textbf{monotonic DP path selection} chooses a globally consistent sequence of spans across the full recording.

\textbf{Search-window determination.}
\label{sec:align-window}
Matching each utterance independently against the entire model word stream is fragile in long conversational recordings: short or common utterances may appear many times and can be matched to the wrong occurrence. We therefore first group consecutive reference utterances into chunks of
$L_{\mathrm{chunk}}$ tokens and align each chunk to the model word stream
$W^{(m)}$ by minimum edit distance, using the resulting matched region as a
coarse search window. Each utterance is then searched within its chunk's matched region, expanded by $\tau_{\mathrm{buf}}$ tokens on each side.

\textbf{In-window candidate search.}
\label{sec:align-candidates}

The search window localizes where an utterance should occur, but a single local
best match can still be unreliable: short utterances, repeated phrases, and ASR errors may create several plausible spans. We therefore treat this stage as
candidate generation rather than final selection. Within its window, each
utterance $u_i$ is matched against the model words by a minimum edit-distance
alignment. This yields the best span ending at each possible word position, and we keep a span $s=[a,b]$ as a candidate only when its word error rate falls below
the threshold,
\begin{equation}
\mathrm{WER}(s) \;:=\; \frac{\mathrm{editdist}(s, u_i)}{|u_i|}
\;\le\; \tau_{\mathrm{wer}},
\label{eq:wer-cand}
\end{equation}
retaining the top-$K$ lowest-WER candidates per utterance. The retained
candidates form the local alternatives passed to the monotonic DP stage, which
then selects a globally order-consistent sequence of spans.

\textbf{Monotonic DP path selection.}
\label{sec:align-dp}

This stage uses a weak turn-order prior: since the model word stream is decoded
from the same recording, it should roughly follow the speaker-turn order of the
reference transcript. This lets us prefer a single left-to-right path through the
candidate spans. The assumption is only approximate, however: ASR may insert
wrong words, skip faint speech, or produce words for speech that is not cleanly
represented in the transcript. We therefore use a soft path cost rather than
forcing every utterance to match.

Given the per-utterance candidate sets, we select one decision per utterance
(one candidate span or \textsc{Missing}) to minimize the path cost, subject to the
chosen spans being monotonic and non-overlapping in $W^{(m)}$. The cost has three
complementary terms: \emph{match cost} measures whether a chosen span is locally
credible, \emph{jump penalty} keeps the sequence in the expected turn order, and
\emph{abstain cost} prevents unreliable utterances from being forced onto
unrelated words.

\emph{Match cost.}
This term is the price of committing to a candidate span. WER measures
token-level mismatch, while the similarity term helps break ties on short or
common utterances by favoring spans whose character sequence resembles the whole
utterance:
\begin{equation}
\ell(s) = \mathrm{WER}(s)
+ w_{\mathrm{sim}}\bigl(1 - \mathrm{sim}(s)\bigr),
\label{eq:localcost}
\end{equation}
where $\mathrm{sim}(s)$ is a character-level sequence-match ratio\footnote{We
use the Ratcliff--Obershelp ratio, a normalized matching-block similarity in
$[0,1]$.} after down-weighting very short or length-mismatched comparisons.
All terms are measured in WER units, so the weights only set their relative
trade-offs.

\emph{Jump penalty.}
This term implements the turn-order prior: neighboring transcript turns should
map to nearby regions of the model word stream. A large unexplained jump usually
indicates that a short or common utterance has matched the wrong occurrence.
With model-word gap $g = a_i - b_{i-1} - 1$ and reference-token gap $r$ between
two consecutive matched spans,
\begin{equation}
\mathrm{jump}(s_{i-1}, s_i)
= \min\!\Bigl(P_{\max},\;
w_{\mathrm{jmp}}\,\max(0,\; g - r - \tau_{\mathrm{jmp}})\Bigr).
\label{eq:jump}
\end{equation}
The cap keeps one bad transition from dominating the whole path.

\emph{Abstain cost.}
This term is the escape hatch for utterances the model does not reliably decode.
It should be cheaper than forcing a poor span, but more expensive than accepting
a credible one; longer utterances are harder to drop because they carry more
lexical evidence:
\begin{equation}
\mathrm{miss}(u_i) = P_{\mathrm{miss}} + p_{\mathrm{tok}}\,\min(L_{\max}, |u_i|).
\label{eq:miss}
\end{equation}

The alignment minimizes the total cost over all valid assignments,
\begin{equation}
  \min \sum_{i:\,\textsc{Matched}}
       \bigl[\ell(s_i) + \mathrm{jump}(s_{i-1}, s_i)\bigr]
     + \sum_{i:\,\textsc{Missing}} \mathrm{miss}(u_i),
  \label{eq:objective}
\end{equation}
which is solved exactly by a Viterbi recursion over the candidate lattice with beam width $B$ (Algorithm~\ref{alg:align}).\footnote{Alignment values used throughout: $L_{\mathrm{chunk}}{=}100$, $\tau_{\mathrm{buf}}{=}5$, $\tau_{\mathrm{wer}}{=}0.75$, $K{=}40$, $w_{\mathrm{sim}}{=}0.5$, $P_{\mathrm{miss}}{=}1.1$, $p_{\mathrm{tok}}{=}0.02$, $L_{\max}{=}20$, $w_{\mathrm{jmp}}{=}0.0188$, $\tau_{\mathrm{jmp}}{=}5$, $P_{\max}{=}3.0$, $B{=}100$.}

\begin{algorithm}[t]
\caption{Per-model alignment}
\label{alg:align}
\begin{algorithmic}[1]
\Require utterances $U$, model word stream $W^{(m)}$, config $\Theta$
\State concatenate consecutive utterances into chunks of at most $L_{\mathrm{chunk}}$ tokens
\For{each chunk $q$}
\State align $q$ to $W^{(m)}$ by minimum edit distance
\State define the buffered chunk window using $\tau_{\mathrm{buf}}$
\For{each utterance $u_i \in q$}
\State find candidate spans within the buffered window
\State keep the top-$K$ candidates satisfying Eq.~\eqref{eq:wer-cand}
\EndFor
\EndFor
\State select a monotonic path with beam $B$
\Comment{Eqs.~\eqref{eq:localcost}--\eqref{eq:miss}}
\State \Return per-utterance span $[a_i,b_i]$ or \textsc{Missing}
\end{algorithmic}
\end{algorithm}

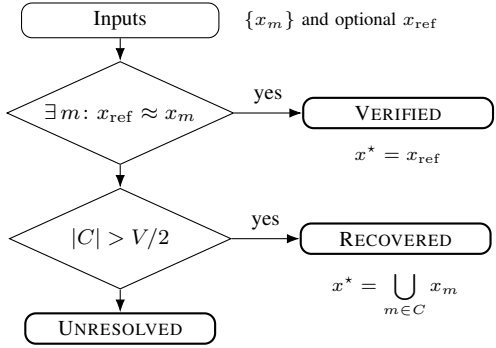
\begin{figure}[t]
\centering
\begin{tikzpicture}[
node distance=3mm and 9mm,
>=Latex,
font=\footnotesize,
box/.style={rounded corners, draw, align=center, inner sep=3pt, text width=24mm},
dec/.style={diamond, aspect=2.2, draw, align=center, inner sep=2pt},
term/.style={rounded corners, draw, thick, align=center, inner sep=3pt, text width=23mm},
ann/.style={font=\scriptsize, align=center}
]

\node[box] (start) {Inputs};
\node[ann, right=2mm of start] (startann) {$\{x_m\}$ and optional $x_{\mathrm{ref}}$};

\node[dec, below=of start] (d1) {$\exists\, m\colon x_{\mathrm{ref}}\approx x_m$};
\node[term, right=of d1] (ver) {\textsc{Verified}};
\node[ann, below=1mm of ver] {$x^\star=x_{\mathrm{ref}}$};

\node[dec, inner sep=4.5pt, below=of d1] (d2) {$|C| > V/2$};
\node[term, right=of d2] (rec) {\textsc{Recovered}};
\node[ann, below=1mm of rec] {$x^\star=\displaystyle\bigcup\limits_{m\in C} x_m$};

\node[term, below=of d2] (unres) {\textsc{Unresolved}};

\draw[->] (start) -- (d1);
\draw[->] (d1) -- node[above]{yes} (ver);
\draw[->] (d1) -- (d2);
\draw[->] (d2) -- node[above]{yes} (rec);
\draw[->] (d2) -- (unres);

\end{tikzpicture}
\caption{Ensemble voting for one utterance. $C$ is the largest block of mutually-agreeing votes and $V$ the number of valid votes.}
\label{fig:voting}
\end{figure}

\subsection{Step 3: Ensemble voting}
\label{sec:ensemble}

Step~2 yields, for each utterance, up to one span per model; these are the votes. The ensemble decides each utterance's timestamp by a majority consensus over those spans, or abstains when the models do not agree. For utterance $u_i$, each model $m$ that produced a match casts a vote: a time interval $x_m = [\alpha_m, \beta_m]$ given by the onset and offset timestamps of its matched word span. When the reference transcript already carries a coarse, often unreliable timestamp for $u_i$, we write that interval as $x_{\mathrm{ref}}$. Two votes $x_m$ and $x_n$ \emph{agree}, written $x_m \approx x_n$, if their endpoints nearly coincide or they overlap substantially with nearby centers,
    \begin{equation}
\begin{aligned}
  x_m \approx x_n \;\;\text{if}\;\; & \max(|\alpha_m{-}\alpha_n|,\,|\beta_m{-}\beta_n|) \le \varepsilon \\
    & \text{or}\;\; \bigl(\mathrm{IoU}(x_m,x_n) \ge \theta \,\land\, |c_m{-}c_n| \le \delta\bigr),
\end{aligned}
\label{eq:close}
\end{equation}
where $c_m$ is the center of span $x_m$, $\mathrm{IoU}(x_m,x_n) = |x_m \cap x_n| / |x_m \cup x_n|$ is the interval intersection-over-union, and $\delta = \max(\varepsilon, \kappa\,\Delta_{\max})$ scales with the longer span's duration $\Delta_{\max}$ through a factor $\kappa$.\footnote{Ensemble values used throughout: $\varepsilon{=}0.5$\,s, $\theta{=}0.9$, $\kappa{=}0.25$.}

We form the largest block $C$ of mutually-agreeing votes\footnote{When non-transitive agreement yields two equally-large blocks, the tie is broken in favor of the block whose votes have higher similarity and lower WER, i.e.\ the larger total quality weight $\sum_{m\in C}\omega_m$ with $\omega_m = \max(0,\sigma_m)\bigl(1-\min(\rho_m,1)\bigr)$ for a vote of similarity $\sigma_m$ and WER $\rho_m$. This tiebreak fires rarely and never applies when a single majority block exists.} and accept it only under a \emph{strict
majority} of the $V$ valid votes, $|C| > V/2$; otherwise the utterance is abstained. The winning block is fused by \emph{union},
\begin{equation}
  [\,\alpha^\star, \beta^\star\,] = \Bigl[\,\min_{m\in C} \alpha_m,\;\; \max_{m\in C} \beta_m\,\Bigr],
  \label{eq:union}
\end{equation}
which becomes the utterance's recovered span. Each utterance is assigned one of three status labels (Figure~\ref{fig:voting}). \textsc{Verified}: the provided span $x_{\mathrm{ref}}$ is corroborated by a model vote and kept. \textsc{Recovered}: no provided span is corroborated (none exists, or $x_{\mathrm{ref}}$ disagrees with the majority), so the fused majority block is used. \textsc{Unresolved}: no reliable majority is found.

\section{Dataset curation}
\label{sec:curation}

We apply \textsc{Beacon} to CHILDES~\cite{macwhinney2000childes}
with the goal of turning long-form child--adult recordings into short,
transcript-aligned speech clips. CHILDES already provides trusted text through
CHAT transcripts, whose speaker-attributed utterance tiers
(\texttt{*CHI:}, \texttt{*MOT:}, \texttt{*INV:}, etc.) define the reference
utterances. However, many recordings have coarse, missing, or unreliable
utterance-level timestamps, which prevents the transcripts from being directly cut into clean clips. Our method recovers more reliable timestamps for these
utterances; the surrounding dataset-curation steps then turn the recovered spans
into a usable release by selecting suitable recordings before alignment and
applying clip-level preprocessing after segmentation.

\begin{table}[t]
\centering
\caption{Statistics of the two released versions (16\,kHz mono WAV, English). Ages use CHILDES \texttt{years;months} notation (e.g.\ \texttt{0;6} is 0 years 6 months). Duration rows give the clip share with audio hours in parentheses; word and vocabulary counts are over verbatim-normalized text.}
\label{tab:datasets}
\begin{tabular}{lrr}
\toprule
 & General & ASR \\
\midrule
\multicolumn{3}{l}{\emph{Coverage}}\\
CHILDES corpora             & 50        & 50 \\
Recordings (sessions)       & 5{,}533   & 5{,}344 \\
Child age range (yr)        & 0;6--14;3 & 0;6--14;3 \\
Utterances (clips)          & 501{,}880 & 310{,}983 \\
Audio (hours)               & 413.3     & 282.6 \\
\midrule
\multicolumn{3}{l}{\emph{Utterance duration}}\\
Mean\,/\,median\,/\,p90 (s) & 2.96\,/\,2.25\,/\,5.1 & 3.27\,/\,2.43\,/\,5.9 \\
$\le 2$\,s                  & 40.0\% (90.8\,h)  & 35.4\% (49.5\,h) \\
2--5\,s                     & 49.8\% (203.6\,h) & 50.7\% (131.9\,h) \\
5--10\,s                    &  8.2\% (76.4\,h)  & 11.1\% (64.1\,h) \\
$>10$\,s                    &  2.0\% (42.4\,h)  &  2.9\% (37.1\,h) \\
\midrule
\multicolumn{3}{l}{\emph{Content}}\\
Words                       & 2.18\,M   & 1.63\,M \\
Vocabulary (types)          & 20{,}581  & 18{,}168 \\
Transcript             & raw CHAT  & verbatim \\
\bottomrule
\end{tabular}
\end{table}

From the recovered timestamps we build two releases in sequence
(Table~\ref{tab:datasets}). The \textbf{general-purpose} child-speech dataset
(Section~\ref{sec:general}) preserves the original CHILDES annotation as much as
possible: clips retain the raw CHAT text, including unintelligibility, pause, and
other annotation markers, so the release is not tied to any single downstream
task. The \textbf{ASR-training} dataset (Section~\ref{sec:asr}) is then derived
as a stricter subset for speech recognition, with verbatim-normalized transcripts
and additional filters for the tight audio-text agreement that ASR training
requires.

\subsection{General-purpose dataset}
\label{sec:general}

\textbf{Selection.}
We start from the English-only CHILDES recordings. A recording-level
quality-control pass drops mislabeled non-English subtrees, effectively silent audio, and recordings that all systems fail to transcribe. We then run the \textsc{Beacon} pipeline and keep only the child (CHI) utterances.

\textbf{Turn merging.}
Very short child utterances (``yeah'', ``mhm'') carry little acoustic context in isolation, which makes their boundaries the hardest to localize and the resulting clips scattered and individually of limited use. We therefore merge temporally adjacent same-speaker (CHI) utterances separated by a gap of at most 1s into a single continuous clip, with 0.5s padding on each side. 
The final merged set is a lossless, \emph{general-purpose} release: all clips retain their raw CHAT text with annotation tags intact.

\subsection{ASR-training dataset}
\label{sec:asr}

For ASR training, the general-purpose release is too permissive: it preserves annotation-rich CHAT text and does not guarantee that every clip is
a clean supervised ASR pair. The ASR release targets a narrower use case: each
clip should be paired with a usable verbatim transcript, and the audio should
contain the speech described by that transcript with little extra or missing
speech. We therefore derive the ASR release as a stricter subset using the following two passes: 

\textbf{Verbatim normalization.}
We build on the Whisper English normalizer\cite{radford2023robust}, but add
CHILDES-specific rules to preserve verbatim child speech while removing CHAT
annotation artifacts.\footnote{The added rules keep fillers such as
\texttt{um}, \texttt{uh}, and \texttt{mhm}; keep filler spellings distinct;
disable contraction expansion and number normalization; strip CHAT markers such
as \texttt{\&=}, \texttt{+//.}, and \texttt{[..]}; and treat
\texttt{xxx}, \texttt{yyy}, and \texttt{www} as exclusion markers.}
After normalization, we filter out clips whose reference text is empty or
contains unintelligible markers.

\textbf{WER-based misalignment filtering.}
Each normalized clip is transcribed by a fixed ASR\footnote{We use
\texttt{Qwen3-ASR-1.7B}, chosen empirically for stable child-speech decoding.}.
We compare the output with the normalized reference and drop the clip if
either the insertion or deletion rate exceeds $0.25$:
$\mathrm{ins}_c/|\mathrm{ref}_c| > 0.25$ or
$\mathrm{del}_c/|\mathrm{ref}_c| > 0.25$. High insertion suggests extra speech
in the audio that is not covered by the reference, while high deletion suggests
reference words that are not supported by the audio. 


\section{Experiments}
\label{sec:experiments}

\subsection{Ablation}
\label{sec:ablation}

\textbf{Setup.}
We validate \textsc{Beacon} on two long-form corpora with reliable utterance timestamps: RSR~\cite{RSR} child speech (132 recordings, 2{,}338 utterances) and AMI meetings~\cite{carletta2005ami} (5 meetings, 2{,}598 utterances). The provided timestamps are withheld, so every utterance is localized from the transcript alone. For the utterances a system labels, we report the mean absolute boundary error (MAE) against the true timestamps.

\textbf{Ensemble vs.\ single models.}
Run alone through the same alignment, the four models reach boundary MAEs of 0.97\,s (Parakeet), 1.65\,s (Qwen3), 2.01\,s (WhisperX), and 3.49\,s (Canary) on RSR; the four-model ensemble reaches 0.44\,s. On AMI the single models reach 1.33, 2.01, 2.06, and 3.81\,s, and the ensemble 1.11\,s.

\textbf{Components.}
Removing one component at a time shows that each contributes. Abstention is what makes labels reliable: without it the pipeline labels almost every utterance, but MAE triples on RSR (0.44 to 1.47\,s) and nearly doubles on AMI. Monotonic DP selection keeps repeated phrases from matching the wrong occurrence: without it, MAE doubles on RSR (0.86\,s) and nearly quadruples on AMI (1.11 to 4.19\,s). Candidate generation lowers RSR MAE from 0.48 to 0.44\,s. The search window keeps long recordings tractable, cutting alignment time 7.4$\times$ on AMI at comparable accuracy.

\subsection{Timestamp quality via the ASR proxy}
\label{sec:results}

\textbf{Setup.}
We evaluate timestamp quality with an ASR-error proxy. This proxy is used only
for relative comparison: under the same fixed ASR system, a better aligned
speech--text pair should generally yield lower WER than a misaligned pair,
because the cut audio contains the speech described by the reference and less
unrelated surrounding speech. 
To avoid bias toward our pipeline, we use another stable ASR system\footnote{\texttt{ibm-granite/granite-speech-4.1-2b-nar}} that's not involved in our pipeline\cite{saon2025granite}.
For each method, we cut and transcribe every candidate clip and report corpus
WER\footnote{For a clip set $\mathcal{C}$, let $r_c=|\mathrm{ref}_c|$ and
$e_c=\mathrm{ins}_c+\mathrm{del}_c+\mathrm{sub}_c$. We report bounded corpus
WER as
$\mathrm{WER}_{\mathrm{corp}}=
\frac{\sum_{c\in\mathcal{C}}\min(e_c,r_c)}
{\sum_{c\in\mathcal{C}} r_c}$.
This is token-weighted, not the mean of per-clip WERs or duration-bin WERs.} We bound the WER, so that a
single hallucinated ASR output cannot dominate the corpus
average. Empty ASR outputs are kept in the metric and counted as fully wrong, rather than discarded, because they
reflect the practical failure mode that the extracted audio does not align with the
reference transcript. We compare against four baselines:

\begin{itemize}
  \item \emph{raw}: the original utterance timestamps shipped in the CHILDES
    \texttt{.cha} files, used as-is;
  \item \emph{BatchAlign2 wav2vec}: TalkBank's official aligner~\cite{BatchAlign2}\footnote{\texttt{https://github.com/TalkBank/batchalign2}} run with its wav2vec2 CTC forced-alignment backend;
  \item \emph{BatchAlign2 whisper\_fa}: the same tool with its Whisper large-v2 forced-alignment backend;
  \item \emph{FASA}~\cite{FASA}: the independent single-model aligner that segments by WhisperX's predicted sentences and keeps only high-confidence reference matches. We run it on the same cached WhisperX transcripts \textsc{Beacon} uses and score the clips it emits.
\end{itemize}

\begin{table}[t]
\centering
\caption{Bounded WER(token-weighted). Hours\,$=$\,retained yield. The right block is each method's distribution of reference tokens (\%) across clip durations. BA-w2v and BA-whsp are batchalign2's wav2vec and Whisper backends.}


\label{tab:timestamp_proxy}
\begin{tabular}{lrrccc}
\toprule
& \multicolumn{2}{c}{Corpus} & \multicolumn{3}{c}{Reference tokens (\%)} \\
\cmidrule(lr){2-3} \cmidrule(lr){4-6}
Method & Hours & WER$\downarrow$ & 0--5\,s & 5--10\,s & $>$10\,s \\
\midrule
raw            & 605.9 & 63.7 & 86 & 11 & 3 \\
\rowcolor{gray!10}
BA-w2v         & 571.6 & 62.3 & 84 & 13 & 3 \\
BA-whsp        & 603.8 & 60.9 & 83 & 14 & 3 \\
\rowcolor{gray!10}
FASA           & 167.6 & 49.7 & 89 & 9 & 2 \\
\midrule
\textsc{Beacon}    & 432.4 & 51.3 & 82 & 14 & 4 \\
\rowcolor{gray!10}
\textbf{$+$ merge} & \textbf{413.3} & \textbf{48.5} & 66 & 22 & 12 \\
\bottomrule
\end{tabular}
\end{table}

\textbf{Results.}
Table~\ref{tab:timestamp_proxy} shows that our fully automatic $+$merge
pipeline gives the best overall timestamp quality under the ASR proxy while
retaining a large portion of the corpus. The improvement comes from two effects:
ensembling complementary ASR systems reduces alignment errors, and turn merging
avoids over-fragmenting short child utterances into tiny clips. The merged clips
provide slightly more acoustic context and shift more tokens into longer
segments, where all methods are more stable.

The baselines reveal two different failure modes. The raw CHILDES timestamps and
the official BatchAlign2 backends remain high-error, suggesting that standard
single-model alignment tools struggle with long, noisy child-speech recordings.
FASA reaches a competitive proxy score, but largely because its design is highly
selective: it keeps only spans that its ASR model already matches with high
confidence. This makes FASA closer to a high-precision, low-yield filter than a
high-recall corpus curator. Because FASA emits standalone ASR segments rather
than speaker-attributed CHAT turns, our same-speaker merging strategy cannot be
reliably applied to it. Overall, our shipped version provides the best trade-off between
proxy accuracy and retained yield for large-scale dataset curation.

\subsection{Downstream usefulness of the curated dataset}
\label{sec:downstream}

\textbf{Setup.}
The ASR proxy evaluates timestamp quality, but it does not by itself show whether
the curated clips are useful for training. We therefore test downstream
usefulness by fine-tuning off-the-shelf ASR models and evaluating them on
held-out child-speech benchmarks. Since FASA reaches a similar proxy accuracy
with a much smaller curated set, we include it as a high-precision, low-yield
baseline. For each model, we compare three settings: zero-shot, fine-tuned on the
FASA-curated set, and fine-tuned on our ASR-training set. Evaluation uses four
children's speech benchmarks held out from training
(RSR~\cite{RSR}, MyST~\cite{MyST}, OGI Kids~\cite{OGI}, and CMU
Kids~\cite{CMU-Kids}), none overlapping the CHILDES data. WER is scored with the
standard Whisper English normalizer applied symmetrically to references and
hypotheses. To check that the effect comes from the \emph{data} rather than one
model family, we repeat the experiment with three off-the-shelf ASR models.

\textbf{Results.}
Table~\ref{tab:downstream} shows that CHILDES-derived curated data is useful for
out-of-domain child ASR: both the FASA-curated set and our ASR-training set
generally improve over the corresponding zero-shot models. This indicates that,
once timestamped and filtered, CHILDES provides transferable child-speech
supervision rather than only corpus-specific signal. The difference lies in yield and
coverage. FASA's selective set is precise enough to help, but its smaller size
limits the training benefit. Our pipeline retains substantially more usable
training audio while maintaining strong timestamp quality, leading to larger
average relative WER reductions across all three model families. Thus the
overall $\Delta$WER summarizes the main trend: both curated sets help, but the
higher-yield set produced by our method gives a stronger downstream gain.

\begin{table}[t]
\centering
\caption{Downstream WER (\%, lower is better) and $\Delta$WER, the mean relative WER reduction over the zero-shot row across the four benchmarks (\%, higher is better). ``$+$\,ours'': fine-tuned on our ASR-training set ($282.6$\,h); ``$+$\,FASA'': fine-tuned on a FASA-curated set ($167.6$\,h).}
\label{tab:downstream}
\begin{tabular}{lrrrrr}
\toprule
& RSR & MyST & OGI & CMU & Avg. $\Delta$WER \\
\midrule
Whisper & 25.2 & 12.8 & 69.3 & 16.3 & -- \\
$+$ FASA (167.6\,h) & 24.9 & 10.8 & 33.9 & 19.3 & $12.4\%\downarrow$ \\
\rowcolor{gray!10}
$+$ ours (282.6\,h) & 21.4 & 11.5 & 24.2 & 18.3 & \textbf{$19.5\%\downarrow$} \\

\midrule

Parakeet & 24.7 & 10.5 & 18.3 & 15.2 & -- \\
$+$ FASA (167.6\,h) & 25.0 & 10.8 & 16.7 & 15.1 & $1.3\%\downarrow$ \\
\rowcolor{gray!10}
$+$ ours (282.6\,h) & 22.5 & 10.7 & 17.1 & 15.1 & \textbf{$3.6\%\downarrow$} \\

\midrule

Canary & 28.9 & \phantom{0}9.5 & 18.6 & 15.8 & -- \\
$+$ FASA (167.6\,h) & 27.0 & \phantom{0}9.4 & 16.1 & 16.2 & $4.6\%\downarrow$ \\
\rowcolor{gray!10}
$+$ ours (282.6\,h) & 26.4 & \phantom{0}9.7 & 15.8 & 15.7 & \textbf{$5.6\%\downarrow$} \\
\bottomrule
\end{tabular}
\end{table}

\section{Limitations}

Our pipeline abstains where the models disagree, trading yield for reliability: it may discard clips whose labels are accurate but whose audio the ASR systems transcribe poorly, and the retained set is biased toward easily recognized speech. The hyperparameters were set from light manual inspection and are likely conservative; since the optimum follows the ASR predictions, which vary with audio condition and speaker, they may need retuning on other corpora.


\section{Conclusion}

We presented \textsc{Beacon}, a fully automatic framework that recovers utterance-level timestamps in long-form recordings by aligning the word timestamps of multiple ASR systems to a trusted transcript and fusing them by consensus, so that model disagreement becomes an abstention signal. Applied to English CHILDES, it yields a general-purpose release that preserves the raw CHAT annotation and an ASR-training subset that improves out-of-domain child ASR across three model families. The recipe applies to any weakly timestamped corpus with a trusted transcript.

\section*{Acknowledgments}

This material is based upon work supported under the AI Research Institutes program by National Science Foundation and the Institute of Education Sciences, U.S. Department of Education, through Award \#2229873 - National AI Institute for Exceptional Education. Any opinions, findings and conclusions or recommendations expressed in this material are those of the author(s) and do not necessarily reflect the views of the National Science Foundation, the Institute of Education Sciences, or the U.S. Department of Education.
This work used the Delta system at the National Center for Supercomputing Applications through allocation bgfd-delta-gpu from the Advanced Cyberinfrastructure Coordination Ecosystem: Services \& Support (ACCESS) program, which is supported by National Science Foundation grants \#2138259, \#2138286, \#2138307, \#2137603, and \#2138296.

\bibliographystyle{IEEEtran}
\bibliography{IEEEabrv,IEEEfull,IEEEexample}

\end{document}